\documentclass{emulateapj}
\usepackage{times}
\usepackage{graphicx}% Include figure files

\renewcommand{\arraystretch}{0.9}

\slugcomment{To apper in {\it The Astrophysical Journal}}

\begin{document}

%\title{{\it Fermi} LAT Performance Illustrated by the Vela Pulsar}
\title{{\it Fermi} LAT Observations of the Vela Pulsar}
\author{
A.~A.~Abdo\altaffilmark{1,2}, 
M.~Ackermann\altaffilmark{3}, 
W.~B.~Atwood\altaffilmark{4}, 
R.~Bagagli\altaffilmark{5}, 
L.~Baldini\altaffilmark{5}, 
J.~Ballet\altaffilmark{6}, 
D.~L.~Band\altaffilmark{7,8}, 
G.~Barbiellini\altaffilmark{9,10}, 
M.~G.~Baring\altaffilmark{11}, 
J.~Bartelt\altaffilmark{3}
% ...
}
\altaffiltext{1}{National Research Council Research Associate}
\altaffiltext{2}{Space Science Division, Naval Research Laboratory, Washington, DC 20375}
\altaffiltext{3}{W. W. Hansen Experimental Physics Laboratory, Kavli Institute for Particle Astrophysics and Cosmology, Department of Physics and Stanford Linear Accelerator Center, Stanford University, Stanford, CA 94305}
\altaffiltext{4}{Santa Cruz Institute for Particle Physics, Department of Physics and Department of Astronomy and Astrophysics, University of California at Santa Cruz, Santa Cruz, CA 95064}
\altaffiltext{5}{Istituto Nazionale di Fisica Nucleare, Sezione di Pisa, I-56127 Pisa, Italy}
\altaffiltext{6}{Laboratoire AIM, CEA-IRFU/CNRS/Universit\'e Paris Diderot, Service d'Astrophysique, CEA Saclay, 91191 Gif sur Yvette, France}
\altaffiltext{7}{Center for Research and Exploration in Space Science and Technology (CRESST), NASA Goddard Space Flight Center, Greenbelt, MD 20771}
\altaffiltext{8}{NASA Goddard Space Flight Center, Greenbelt, MD 20771}
\altaffiltext{9}{Istituto Nazionale di Fisica Nucleare, Sezione di Trieste, I-34127 Trieste, Italy}
\altaffiltext{10}{Dipartimento di Fisica, Universit\`a di Trieste, I-34127 Trieste, Italy}
\altaffiltext{11}{Rice University, Department of Physics and Astronomy, MS-108, P. O. Box 1892, Houston, TX 77251, USA}
% ...

\begin{abstract}

	The Vela pulsar is the brightest persistent source in the GeV
sky and thus is the traditional first target for new $\gamma$-ray 
observatories. We report here on initial {\it Fermi} Large Area Telescope
observations during verification phase pointed exposure and early sky survey scanning.
We have used the Vela signal to verify {\it Fermi} timing and angular resolution. 
The high quality pulse profile, with some 32,400 pulsed photons at 
$E\ge 0.03$\,GeV, shows new features, including
pulse structure as fine as 0.3\,ms and a distinct third peak, which
shifts in phase with energy. 
We examine the high energy behavior of the pulsed emission; initial
spectra suggest a phase-averaged power law index of 
$\Gamma=1.51^{+0.05}_{-0.04}$ with an exponential cut-off at 
$E_c=2.9\pm 0.1$\,GeV.
Spectral fits with generalized cut-offs of the form $e^{-(E/E_c)^b}$ require
$b\le 1$, which is inconsistent with magnetic pair attenuation, and
thus favor outer magnetosphere emission models.
Finally, we report on upper limits
to any unpulsed component, as might be associated with a surrounding
synchrotron wind nebula (PWN). 
\end{abstract}


\section{Introduction}

Radio pulsations at $P$=89\,ms from PSR B0833$-$45 (=PSR J0835$-$4510) 
in the Vela supernova remnant
were discovered by Large, Vaughan \& Mills (1968). This pulsar is bright 
($S_{1.4}\,{\rm GHz} \approx 1.5$\,Jy), young (characteristic age 
$\tau_c = P/2{\dot P} = 11$\,kyr),
and energetic (${\dot E}_{SD}=6.9\times 10^{36}I_{45}\, {\rm ergs/s}$,
for a neutron star moment of inertia $10^{45}I_{45}\,{\rm g\,cm^2}$).
It is embedded in a flat spectrum radio synchrotron nebula (see Frail et al.
1997) and is surrounded by a bright X-ray
wind nebula displaying remarkable toroidal symmetry
(Helfand et al. 2001; Pavlov et al. 2003).
A VLBI parallax measurement provides a well-determined distance of $D=287_{-17}^{+19}$pc 
(Dodson et al. 2003), improving on earlier optical HST measurements
(Caraveo et al. 2001). This proximity means that the spindown energy flux at Earth,
${\dot E}_{SD}/4\pi d^2$, is second only to that of the Crab pulsar and ensures 
that Vela is among the most intensely studied neutron stars, particularly
at high energies.

	Pulsed $\gamma$-ray emission from Vela was first detected during
the {\it SAS-2} mission (Thompson et al. 1975); it is, in fact, the brightest 
persistent source of celestial $\gamma$-rays and has been the proving ground 
of GeV observatories ever since.  The basic source properties
discovered by {\it SAS-2} were elaborated using observations by {\it COS-B} 
(Kanbach et al. 1980), and {\it CGRO}/EGRET (Kanbach et al. 1994): the 
source is $\sim$100\% pulsed with no convincing
evidence for year-scale variations, the two $\gamma$-ray peaks are separated 
by 0.42 in pulsar phase and the first peak lags the radio peak
by 0.12 in phase. The source spectrum is
hard with an average photon index $\Gamma= 1.7$ with evidence of a cut-off
above 2--4\,GeV. The most detailed study, to date, of the $\gamma$-ray pulsations is
that of Fierro et al. (1998), which produced a high signal-to-noise pulse profile 
and evidence for phase-resolved spectral variations. As the present work 
was being prepared for submission, the AGILE team reported their initial 
results on $\gamma$-ray pulsars, including Vela (Pellizoni et al. 2008); the results, 
given the more limited count rates and energy resolution, are broadly consistent 
with our conclusions described below.

With the successful launch of the {\it Fermi} Gamma-ray Space Telescope,
formerly GLAST, observatory on June 11, 2008,
we have a new opportunity to examine the high energy behavior of the Vela
pulsar and to study this archetype of the young, energetic pulsars in detail. During
Launch and Early Operations phase (L\&EO) the {\it Fermi} mission targeted the Vela pulsar
for a number of pointed observations, in addition to coverage during 
initial tests of the sky survey mode. In the latter mode the instrument axis is
offset North and South of the zenith during alternate orbits to provide 
near-uniform sky coverage 
every three hours.  One main purpose of these early observations was to tune the
Large Area Telescope (LAT) performance on celestial $\gamma$-ray sources. However,
the initial results on Vela itself are of interest, including new 
high energy features in the pulse profile, an improved measurement
of the high energy cut-off in the pulsar spectrum, and a search for associated
pulsar wind nebula (PWN) emission at GeV energies.

%%%%%%%%%%%%%%%%%%%%%%%%%%%%%%%%%%%%%%%%%%%%%%%%%%%%%%5
\placetable{1}
% To be fair not really sure that we need this....
\begin{table*}
\centering
\caption{\label{1} Early LAT observations of PSR B0833$-$45}
%\bigskip
\begin{tabular}{lcccc}
\hline\hline
Date       & MJD &Primary mode& $N$(E$>$0.03GeV)$^a$ & Notes\\
\hline
%removed - no photons
%2008 Jun 25--Jun 30&54642.73--54647.38& Survey+Pointings &18&Initial LAT calibrations.
%\\
2008 Jun 30--Jul 4&54647.40--54651.36& Sky survey &1859&LAT First light.\\
2008 Jul 4--Jul 15&54651.38--54662.10& Survey+Pointings &5172&LAT 
Calibrations.\\
2008 Jul 15--Jul 19&54662.12--54666.08& Pointings+Survey+limb Following 
&1170&Pointed Obs. Tuning\\
2008 Jul 19--Jul 22*$^b$ &54666.10--54669.14& Pointed Vela+ 2$^{nd}$ 
Target &7751&Pointed Obs. Tuning\\
2008 Jul 22--Jul 24&54669.16--54671.41& Pointed Vela+ 2$^{nd}$ Target 
&3212&Pointed Obs. Tuning\\
2008 Jul 24--Jul 30*&54671.36--54677.45& Pointed Vela+ 2$^{nd}$ Target 
&7607&Pointed Obs. Tuning\\
2008 Jul 30--Aug 3&54677.45--54681.66& Polar study + Sky Survey 
Tuning&886&Nominal Ops.\\
2008 Aug 3--Sep 14*&54687.68--54723.91&Sky Survey&26680&Nominal 
Ops.$^c$ \\
\hline
\multicolumn{5}{l}{$^a$All photons (pulsed and background) within 5$^\circ$ of PSR B0833-45, 
zenith angle $z<$105$^\circ$, event class \emph{Pass6-Diffuse}.}\\
\multicolumn{5}{l}{$^b$Observations marked by asterisk are those used for spectral studies.}\\
\multicolumn{5}{l}{$^c$L$\&$EO ended on Aug.11.}
\end{tabular}
\end{table*}
%%%%%%%%%%%%%%%%%%%%%%%%%%%%%%%%%%%%%%%%%%%%%%%%%%%%%%5
\begin{figure*}
\includegraphics[scale=0.90]{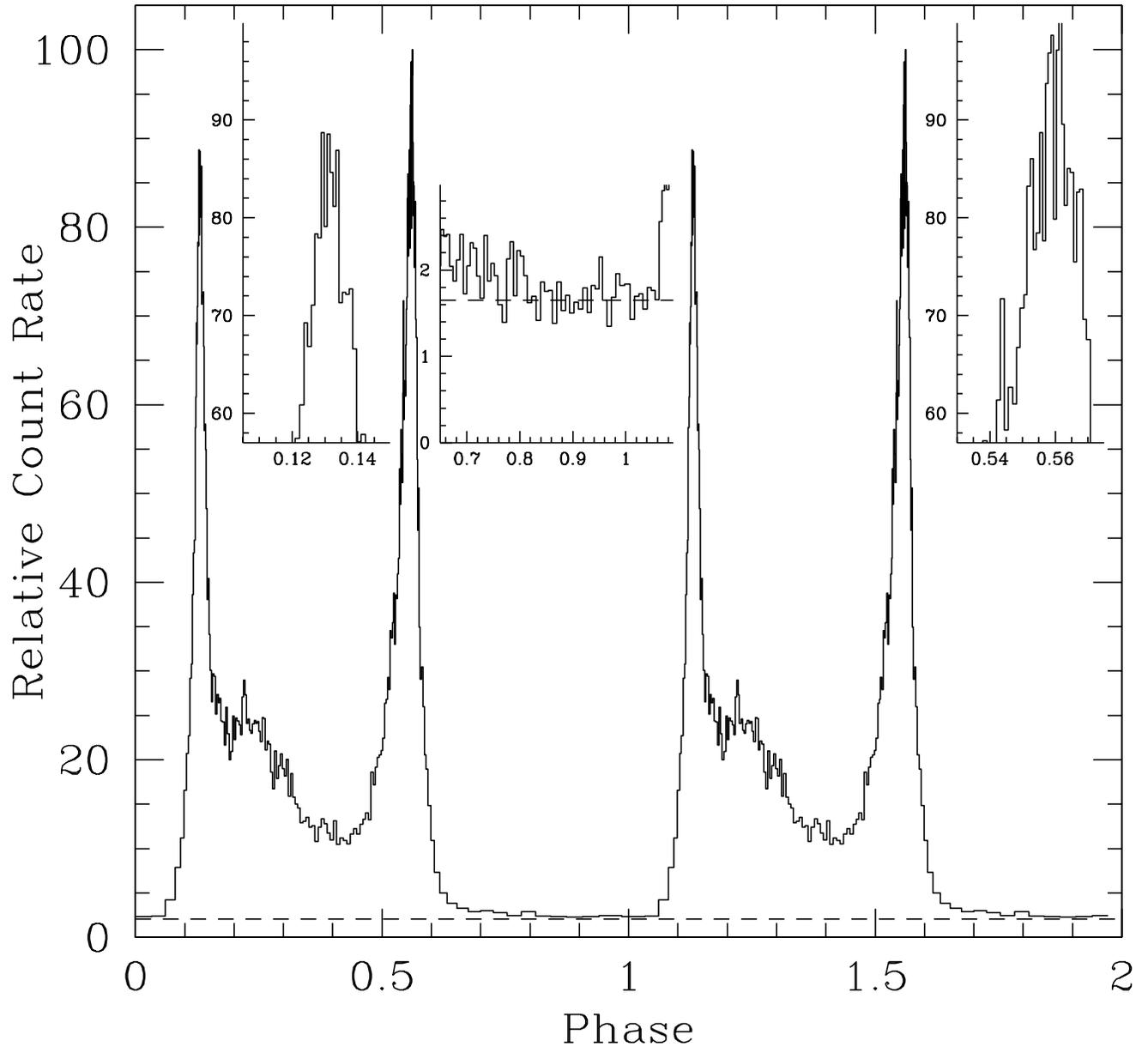}% converted from .eps with epstopdf
\caption{\label{fig:Lightcurves} Vela broad band ($E=0.1-10$\,GeV) pulse profile 
for all photons from an energy dependent ROI. Two pulse periods are shown. The peak
of the radio pulse is at phase $\phi=0$. The count rate is shown in variable width
phase bins with a constant number of counts.
The dashed line shows the background level, as estimated
from a surrounding annulus during the off-pulse phase. Insets show the pulse shape near
the peaks and in the off-pulse region.}
\end{figure*}

\section{Radio Timing Observations} 

The Vela pulsar is young and exhibits substantial timing irregularities.
This means that the optimal use of the Vela $\gamma$-ray photons, which arrive
at an orbit-averaged rate of one every 4 minutes during LAT sky survey observations, 
requires a simultaneous radio ephemeris.
The radio ephemeris is obtained using observations made with the
64-m Parkes radio telescope as part of the overall program for
pulsar monitoring in support of the {\it Fermi} mission (Smith et al. 2008).
A total of 27 times-of-arrival (TOAs) were obtained at a frequency of 1.4\,GHz over
the period of the LAT observations. The median error in each TOA
is 1.7\,$\mu$s. Fits to the TOAs were made with the pulsar's rotational frequency
and frequency derivative as free parameters and the ephemeris created
from these data maintains phase with an rms residual of 90\,$\mu$s, or
$10^{-3}$ of pulsar phase, throughout the LAT observations.
No large timing irregularities or glitches were detected.
An accurate determination of the pulsar's dispersion
measure of $67.95\pm0.03\,{\rm cm^{-3}pc}$ was obtained by measuring the
delay in the TOAs of the pulse across the 256 MHz bandpass of the 1.4 GHz receiver.
This allows extrapolation of the radio ephemeris to infinite 
frequency with an error of $\sim$60\,$\mu$s.
Photon arrival times were referred to the Solar-System barycenter and pulse phases were 
assigned using the standard pulsar timing software TEMPO2 (Hobbs et al. 2006).

	We thus can assign a pulsar phase to the $\gamma$-ray photons referenced to
the radio with high confidence. Gamma-ray events recorded with the LAT have timestamps
that derive from a GPS clock on the {\it Fermi} satellite. Pre-launch ground timing measurements
using cosmic muons seen coincidentally in the LAT and in an independent counting
system demonstrated that the LAT measures event times relative to the spacecraft clock 
with a resolution of better than 300\,ns.  On orbit, satellite telemetry 
indicates a comparable resolution.  The contribution to the barycenter times
from uncertainty in the LAT's position is negligible.  It is also useful to confirm
the absolute timing using celestial signals; as of this writing the six classical
pulsars detected by EGRET have provided good quality LAT pulse profiles. The
phasing of the radio and $\gamma$-ray peaks is consistent with the best EGRET results,
ensuring absolute timing relative to the radio time system better than $\sim 1$ms.

\section{Gamma-Ray Observations}

The LAT instrument on {\it Fermi} is described by Atwood et al. (2008) and references
therein. The LAT is an electron-positron pair production telescope featuring
solid state silicon trackers sensitive to photons from $<30$\,MeV to $>300$\,GeV. It
has a large $\sim$2.4sr field of view, and compared to earlier $\gamma$-ray missions, 
has a large effective area ($\sim 8000\,{\rm cm^2}$ on axis), improved resolution 
($\sim 0.5^\circ$, 68\% containment radius at 1\,GeV for events collected in the
`front' section with thin radiator foils) and small dead time 
($\sim 25\,\mu$s/event).  We report here
on the LAT's initial observations of the Vela pulsar, using data collected during 
35 days of on-orbit verification tests and the initial $\sim 40$ days 
of the on-going first year sky survey.
These data already suffice to illustrate the power of the LAT for astronomical
observations and, indeed, show several new features in the radiation of this
well-known $\gamma$-ray pulsar.

Table 1 contains a journal of the Vela coverage, along with numbers of 
photons with measured energy above $0.03$\,GeV recorded within a radius 
of 5$^\circ$ of the pulsar, including background photons. For both
pointed and survey observations we exclude events within $8^\circ$
of the Earth's limb to minimize contamination by `Earth albedo' photons. 

During the L\&EO period, the instrument configuration
was being tuned for optimum performance. Accordingly, the knowledge of the energy
scale and effective area are more limited than for routine operations.
We discuss how these data verify the LAT photon selection, effective 
area, timing, photon energy measurement and the variation of the point spread 
function (PSF) with energy. These
results may be of use to researchers seeking to predict {\it Fermi} capabilities
during longer exposures of fainter sources, in both sky survey and pointed mode.

\section{Results}

\subsection{Pulse Profiles}

After calibration of the LAT pointing axis on a large set of identified
high latitude $\gamma$-ray sources, we find that the best fit position 
of the Vela point source is within 0.5$^\prime$ of the radio pulsar 
position with a statistical error of 0.4$^\prime$(95\%); the LAT
has unprecedented accuracy for localization of bright hard 
$\gamma$-ray sources, although some systematic uncertainties remain.

The Vela pulsar is embedded in the bright $\gamma$-ray 
emission of the Galactic plane, which is particularly strong at low energies. 
Further, the LAT, like all pair
production telescopes, has an angular resolution dominated by scattering at low
photon energies $\theta_{68} \approx 0.8^\circ E_{GeV}^{-0.75}$ (see below). 
Thus selection of photons from an energy-dependent region of interest (ROI) 
around Vela is important; the best selection depends on the desired product. Here, 
we seek pulse profiles with good signal to noise over a broad energy range, so 
we select photons within an angle 
$\theta<{\rm Max}[1.6-3{\rm Log_{10}}(E_{\rm GeV}),1.3]$\,deg of the pulsar position. 
This includes
a larger fraction of the PSF at high energies, where the background is relatively
faint.  We use `Diffuse' class events, those reconstructed
events having the highest probability of being photons.

In this energy-dependent aperture we have 32,400$\pm$242 pulsed photons and 
$2780\pm53$ background photons with measured energy $>0.03$\,GeV.
Figure 1 shows the 0.1-10\,GeV pulse profile from this ROI with the peak of 
the radio pulsar signal at phase 0.  To highlight the fine structure 
we plot in Figure 1 the pulse profile using variable-width bins, each containing
200 counts. These counts, divided by the bin width, give the photon
flux in each phase interval; these phase bin fluxes thus have a $1\sigma$ 
Poisson statistical error of 7\%. The pulse profile is normalized to 100 at the
pulse peak. Three insets show structure near the first peak (P1), second peak (P2)
and `off-pulse' window at a finer scale. These 
peaks are at pulsar phases $\phi=0.130\pm 0.001$ and $\phi=0.562\pm0.002$, 
respectively.

	The peaks are asymmetric. In particular P2 has a slow rise
and a fast fall. P1 also has a steep outer edge with the fall
somewhat slower. We fit each peak in Figure 1 with two half-Lorentzian 
functions (with different widths for the leading and trailing sides) over
phase intervals which avoided the complex bridge flux
($0.11<\phi<0.16$ for P1, $0.53<\phi<0.59$ for P2).  The outer edges of the
two peaks had consistent Lorentzian half-widths of $\phi = 0.012\pm0.001$. 
Extra structure at the pulse peaks made the half maximum widths somewhat
smaller ($\phi = 0.009\pm 0.002$; 0.8\,ms). The falling edge of peak 1
has a Lorentzian half width of $\phi = 0.017\pm0.0015$, while the 
rising edge of P2 has a width $\phi = 0.027\pm0.005$.
Both peaks show apparently significant structure on scales as small as
$\delta \phi \approx 0.003$ ($\sim 300\, \mu$s), but additional counts are
required to fully probe the pulse profile at this scale.
We also show an inset of the `off pulse region'
($\phi =0.65 - 1.05$), with 50 counts/bin
to show fine structure. Here and in the main panel, the dashed line shows
the estimated residual unpulsed background counts in our energy-dependent ROI,
as measured from the pulse minimum. This is in good agreement with a
background level estimated from $3-5^\circ$ around the pulsar.
We see that, as for other non-radio Vela pulse profiles, there is a faint tail 
of pulsed emission in this region, reaching non-detectability only 
near $\phi = 0.8-1.0$. We estimate that this `off pulse' window contains
235$\pm$15 pulsed photons ($7.3 \times 10^{-3}$ of the pulsed flux).

\begin{figure*}[t!]
\includegraphics[scale=0.8]{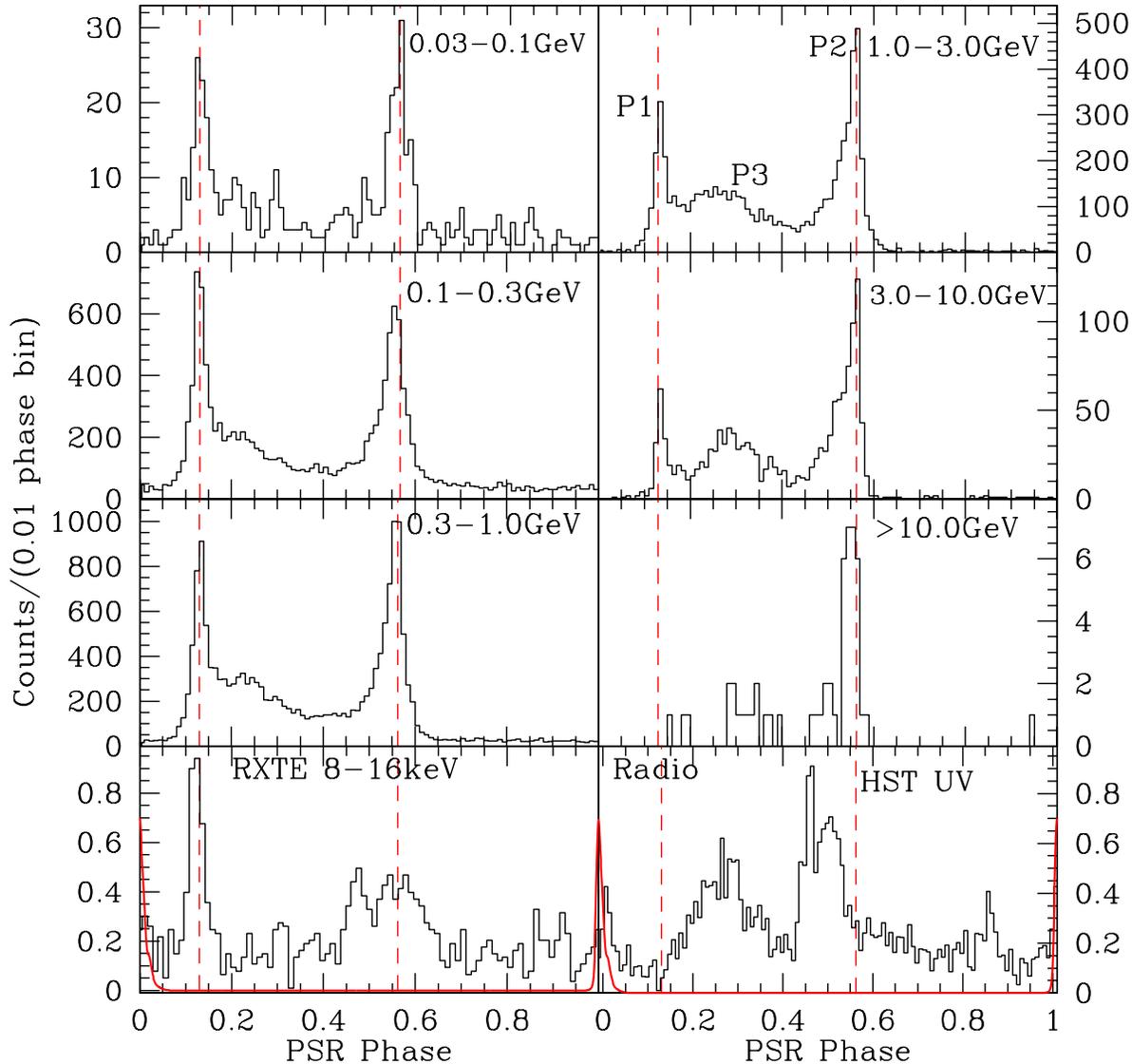}% Here is how to import EPS art
\caption{\label{fig:6bands}The evolution of the Vela $\gamma$-ray pulse profile 
over three decades of energy. Each pulse profile is binned to 0.01 of pulsar phase,
and dashed lines show the phases of the P1 and P2 peaks determined from the 
broad band light curve.
In the top right panel we label the main peaks P1, P2 and P3.
In the bottom panels we show at left the 8-16\,keV {\it RXTE} pulse profile of 
Harding et al. (2002) along with the radio pulse profile (in red).  At lower right,
the 4.1-6.5\,eV {\it HST}/STIS NUV pulse profile of Romani et al.\ (2005) is shown 
for comparison.}
\end{figure*}

\begin{figure}
\includegraphics[scale=0.45]{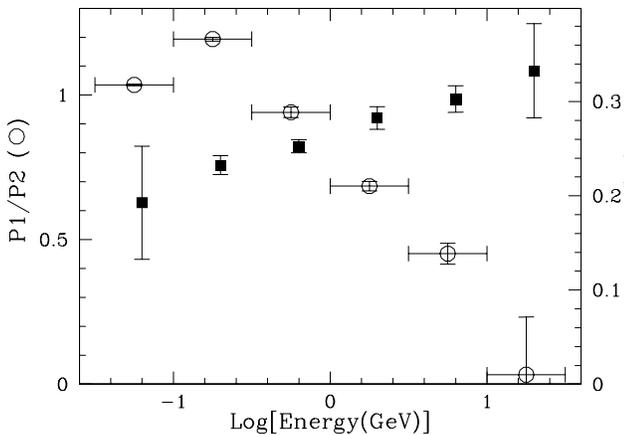}% converted from .eps with epstopdf
\caption{\label{fig:peaks} The evolution of the P1/P2 pulse peak ratio (open circles)
and the P3 pulse peak phase (solid squares) with energy.}
\end{figure}

	There is dramatic energy evolution in the $\gamma$-ray
pulse profile. Figure 2 shows pulse profiles from six energy bins, covering
three decades of the LAT data, drawn from our energy-dependent aperture.
Note that the relatively large aperture at low energies leads to larger
off-pulse flux from the surrounding background. We should also caution that,
with the present limited accuracy in energy measurement below 0.1\,GeV, some
higher energy photons may leak into the lowest energy bin.
The most prominent pulse feature is a decrease
of P1 relative to P2 with increasing energy; P1 is not detectable
above $\sim$10\,GeV (Figure 2). This confirms a trend seen in the EGRET data
for the Crab, Vela and Geminga pulsars -- the second pulse dominates
at the highest energies (Thompson 2001). Interestingly, at the lower energies, below
$\sim 120$\,MeV, the trend is reversed with P1 weakening again with 
respect to P2. We do not find any statistically significant evidence
for shifts in the phases of the narrow P1 and P2 pulse components with
energy. These structures, spanning together $<0.07$ of the rotational phase,
contain emission dominating over three decades of the pulsar photon output.

The `bridge' region between the peaks also shows
appreciable structure with a trailing shoulder of P1 shifting
to later phase. At $>$1\,GeV it is clear that this is a distinct
pulse component (P3) with a peak at $\phi \sim 0.27$ in the 3-10\,GeV
band (Figure 2). Note that this peak shifts in phase by $\delta \phi
\approx 0.14$ between 0.2 and 15\,GeV (Figure 3).

\begin{figure*}[ht!!]
\begin{center}
\includegraphics[scale=0.77]{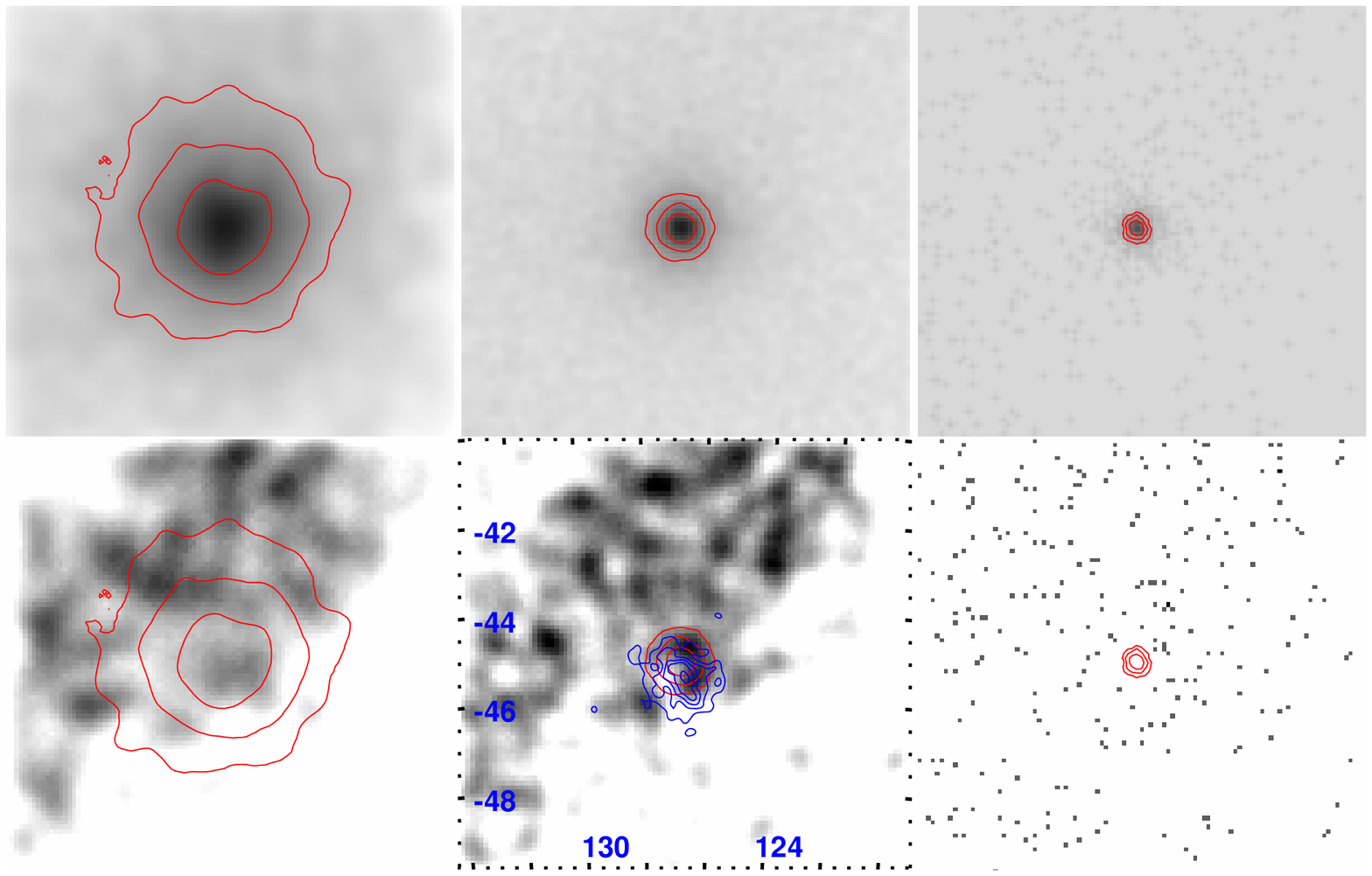}% Here is how to import EPS art

\includegraphics[scale=0.80]{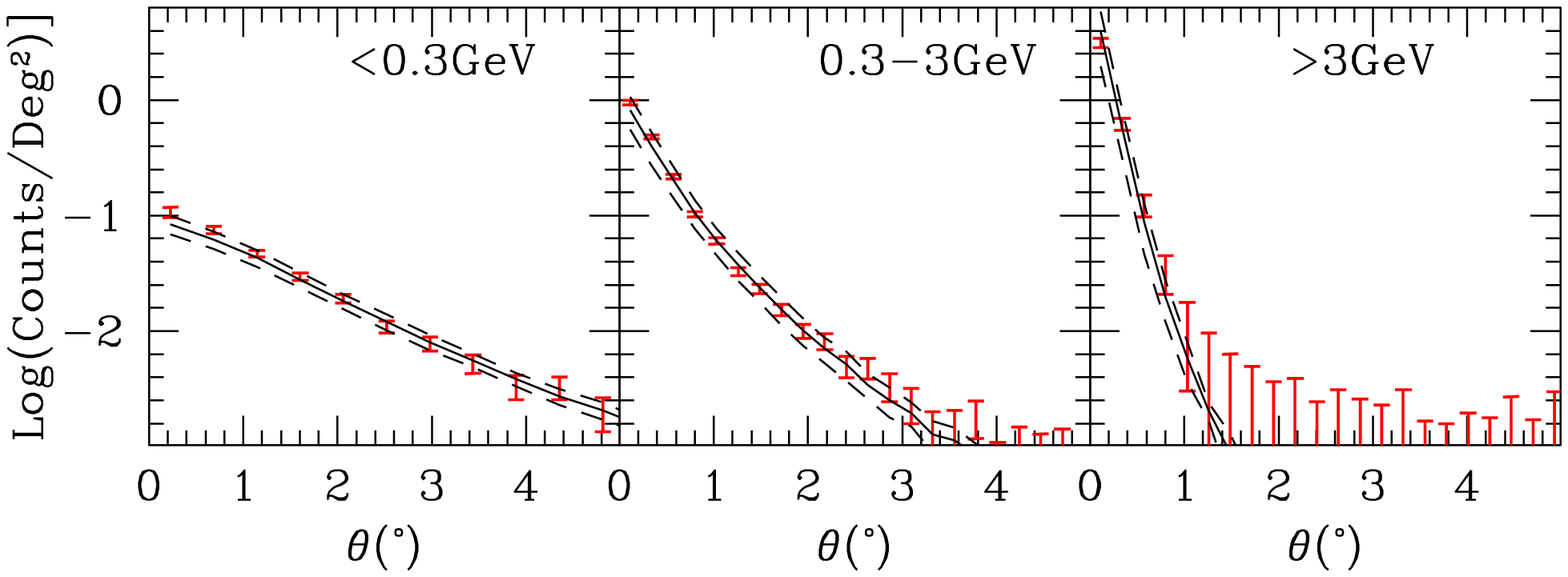}% Here is how to import EPS art
\end{center}
\caption{\label{fig:onoff}On- ($\phi=0.05-0.65$, top row) and 
off- ($\phi=0.65-1.05$, middle row) pulse phase images
in three energy bands. Each panel spans $10^\circ \times 10^\circ$ in equatorial 
coordinates centered on Vela (RA and Dec marked in degrees in the 
central panel). On-pulse data are dominated by the point source; the panels have a hard
square root stretch to show the faint wings of the PSF.  The off-pulse images (linear
grey scale) are dominated by Galactic diffuse emission; some of the structure is due
to limited count statistics. Contours of the
point source at 0.5, 0.25 and 0.125 of the peak intensity are shown (red) in each frame. 
In the middle energy off-pulse image we also show the HESS Vela PWN contours
(Aharonian et al.\ 2006; blue).
Along the bottom row background-subtracted Vela PSFs (red error bars give the statistical
error in the observed counts) are compared with 
simulations based on pre-launch descriptions of expected LAT performance
(solid lines) and estimates of the systematic uncertainty in the model PSF
(dashed lines).}
\end{figure*}

	To compare with the lower energy structure, we show at bottom left
of Figure 2 the 8-16\,keV non-thermal X-ray pulse measured by {\it RXTE} 
(Harding et al.\ 2002).  On the right we similarly show
the 4.1-6.5\,eV NUV {\it HST} STIS/MAMA pulse profile of Romani et al.\ (2005).
The 1.4\,GHz radio pulse profile, whose peak defines phase $\phi=0$, 
runs across the bottom two panels. Both the optical/UV and the hard
X-rays are dominated by non-thermal magnetospheric emission.
In contrast, the $<1$\,keV soft X-ray flux is dominated by thermal emission
from the neutron star surface (e.g. Manzali, de Luca \& Caraveo 2008), making it 
much more difficult to trace the non-thermal peaks through this
intermediate energy band.

We may compare components in the UV and hard X-ray pulse profiles shown with those of
the non-thermal $\gamma$-ray pulsations. In particular,
we should note that the $\gamma$-ray P1 component is dominant in the non-thermal 
X-rays. This component is absent in the optical/UV, but pulse profiles at these energies
have a strong peak in the bridge region at $\phi \sim 0.25$, well matched in phase to the
P3 structure in the $>$\,GeV pulse profiles. This possible connection between 
optical pulses and the $\gamma$-ray bridge emission was already
noted by Kanbach et al.\ (1980).  Note that the profile for UV energies has a distinct
sharp pulse coincident with the radio at $\phi = 0$. While the general
slope of the faint $\phi = 0.65-1.05$ $\gamma$-ray emission matches
that of the UV, no sharp $\gamma$-ray pulse components are yet visible 
in this phase interval.

\begin{figure*}[h!]
\includegraphics[scale=0.9]{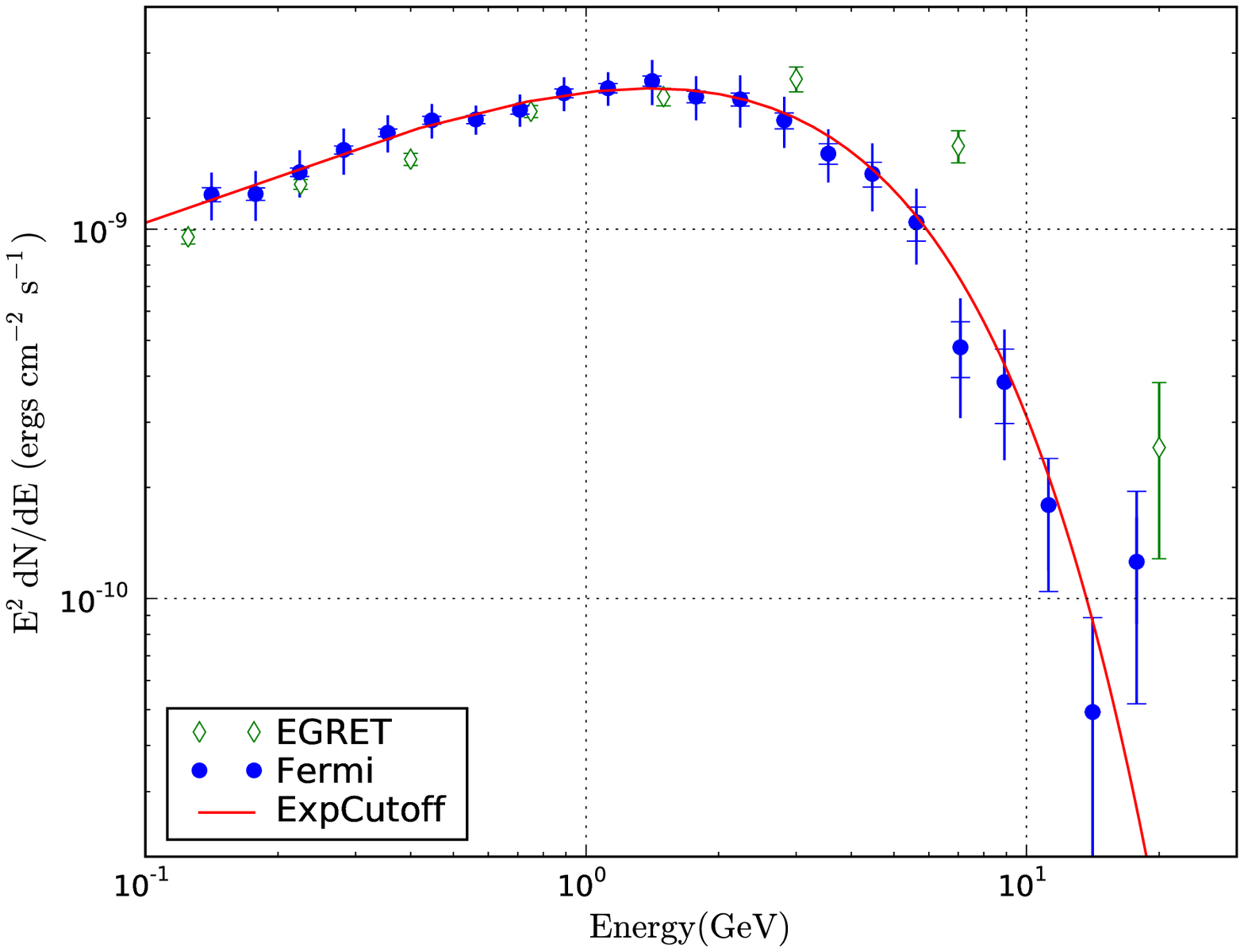}% converted from .png with convert
\caption{\label{fig:spectrum} The phase-averaged Vela spectral energy distribution
($E^2{\rm d}N_\gamma/{\rm d}E$). Both statistical
(capped) and systematic (uncapped) errors are shown. We believe that the
latter are conservative; they dominate at all energies below 7\,GeV.
EGRET data points (diamonds, Kanbach et al.\ 1994) are shown for comparison. 
The curve is the best-fit power law with a simple exponential cut-off.}
\end{figure*}

\subsection{The PSF and a Search for Unpulsed Emission}

	Figure 1 makes it clear that the Vela extraction region is strongly
dominated by the pulsed emission. However there are substantial phase windows
where the pulsar is very faint.  Figure 4 presents {\it Fermi} $\gamma$-ray 
images of the Vela pulsar in three energy bands in the LAT range separated 
into the on- ($\phi=0.05 - 0.65$) and off- ($\phi= 0.65-1.05$) pulse phase regions.
The images (except for 
the low count high energy off-pulse frame) are smoothed by Gaussians
(with $FWHM=1.2^\circ$, $0.7^\circ$ and $0.2^\circ$ for the low 
to high energy images, respectively).
The off-pulse images show the bright diagonal band of the Galactic
diffuse emission, with appreciable structure. Comparison with the 
PSF contours (red) show that this is resolved in the vicinity of
Vela at the higher energies. While a clump of extended emission surrounds
the pulsar position on the few degree scale, it is at present not possible
to associate this flux with known Vela PWN structures (e.g.
the Vela-X TeV PWN, contours in the middle energy band).

	We have subtracted these off- pulse images (scaled by $1.5\times$
to normalize exposure) from the on- pulse images and plotted (error bars) 
the profile of the resulting point source (bottom panels). 
The lines show simulations of the expected PSF 
from pre-launch calibrations (`Pass6'), computed for a source with a 
Vela-like spectrum and off-axis angles similar to the actual observations.
The agreement is quite good. 

	As a first attempt at constraining unpulsed (e.g. PWN)
emission from Vela we have attempted to fit for a point source in 
the off-pulse phase window, fixed at the position of the pulsar.  Using 0.1--10\,GeV
photons in an $8^\circ$ ROI in the $\phi=0.65-1.05$ phase interval,
we derive a 95\% CL upper limit on the flux of $1.8\times 10^{-7}
\gamma/{\rm cm^2/s}$. After subtracting the estimated remnant
pulsed flux in this window (0.73\% of the phase-averaged flux) and scaling
to the full pulse phase, this provides a limit on an unpulsed point
source at the position of Vela of 2.8\% of the 
$E>100$\,MeV pulsed emission count-rate. More photons, especially at higher
energy, will be required to search for a resolved PWN correlated with the
TeV or radio structures.

\subsection{Energy Spectrum}

	To study the phase-averaged spectrum of Vela, we used the standard
maximum-likelihood spectral estimator `gtlike' to be provided with the 
{\it Fermi} SSC science tools.  This fits a source model to the data, 
along with models for the isotropic (instrumental and extragalactic) and 
structured Galactic backgrounds. The instrumental background was comparable
to estimates from pre-launch simulation (Atwood et al.\ 2008).  We used data from the
observation spans indicated in Table 1, selecting photons with 
$E>0.1$\,GeV within 15$^\circ$ of Vela.  Our basic model for the spectrum 
of Vela is a simple power-law with an exponential cut-off.

With the large number of events collected for Vela, the statistical
errors are very small.  Systematic errors for the LAT are still
under investigation (Abdo et al.\ 2009).  %<==(Bruel et al. 2008) 
Here we adopt conservative estimates of the systematic uncertainty in the LAT
effective area, derived from the on-orbit estimation of the photon selection 
efficiency as function of energy and off-axis angle. This estimation 
has been made by comparing Vela on-pulse event selection to event
selection in the off-phase bins. This allows an estimate of the
efficiency for a pure photon sample, which is compared with our current 
simulation of the LAT response. We conducted this exercise for a range of energy
and off-axis angle bins, and use the precision of the resulting effective
area estimates and the difference from the simulated effective area as 
an estimate for the uncertainty in the LAT effective
area as a function of incident photon energy.  This varies from $<10\%$ near 1 GeV to
as much as 20\% for energies below 0.1 GeV and 30\% for energies greater than 10 GeV.

	The gtlike fit for $0.1<E<30$\,GeV is unbinned, and results in a spectrum
of the form
\begin{equation}
{{{\rm d}N}\over{{\rm d}E}} = (2.08\pm0.04\pm0.13) \times 10^{-6}
E^{\Gamma}e^{-E/E_c}  \gamma/{\rm cm^2/s/GeV},
\end{equation}
with $E$ in GeV, $\Gamma=-1.51\pm0.01\pm0.07$ and $E_c=2.857\pm0.089\pm0.17$\,GeV.
The first errors are the statistical values for the fit parameters, while the
second errors are our propagated systematic uncertainties. The latter are
particularly large for our spectral index parameter $\Gamma$ because of the
large uncertainty at present in the effective area at high and low energies.
To explore the stability of this result, we also fit using three other spectral estimators.
We used a standard XSPEC analysis with our best model response matrices, a
binned maximum likelihood estimator which computes the on-pulse
photon counts in a point source weighted aperture in excess of 
off-pulse background counts (`ptlike') and a method which propagates
the model spectrum through simulated instrument response to 
compare with observed pulsed source counts.
For additional tests of the stability of the spectral fits,
the data were also analyzed separately for `pointed' and `survey' observations,
as well as for events in the front (thin radiator foil) and back 
(thick radiator foil) sections of the LAT detector. The
fit parameters for the various data sets and analysis techniques, do
have some statistically significant variation, but all 
were well within our presently-estimated systematic errors, as listed in Equation (1). 
We believe that as our understanding of the LAT performance improves we should be
able to substantially decrease these systematic uncertainties.

	Figure 5 presents the spectral power $E^2{{\rm d}N\over{{\rm d}E}}$
along with this best fit model. The binned spectral points are drawn from
the ptlike analysis and show both the statistical error flags 
(bars with caps) and the inferred systematic error flags (bars without caps); 
the systematic errors dominate for all energies below 7 GeV. We also plot 
the EGRET data points of Kanbach et al.\ (1994). These are discrepant with the LAT points,
particularly above a few GeV. Some studies have indicated that the EGRET
response was incorrectly estimated (Baughman 2007, Stecker et al 2008).

	We have also attempted to fit the data with a generalized cut-off model
of the form Exp[$-(E/E_c)^b$]. We find $b = 0.88 \pm 0.04_{-0.52}^{+0.24}$ 
so that models with a hyper-exponential behavior are well
excluded. Taking into account the systematic errors the spectrum
is fully consistent with the simple exponential $b=1$ cut-off. To check the exclusion of
a given hyper-exponential model we compute the probability of incorrect
rejection of a given $b$ value using the likelihood ratio test. For
example, if only statistical errors are included $b=2$ is rejected at
16.5$\sigma$; inflating the errors to the level of our estimated
systematic uncertainty leads to a small, but not negligible,
0.29\% chance of incorrectly rejecting the $b=2$ model.

The analyses of Kanbach et al.\ (1980), Grenier et al.\ (1988) and Fierro et al.\ (1995) 
already showed that there is significant spectral variation
through the pulse, with the bridge region harder than the peaks. Clearly, in Figure
2 we see that the situation is even more complicated, with three or more overlapping
physical components contributing to the energy-dependent pulse profiles. 
Also, in Figure 5, one sees that there may be emission in the highest energy bins
in excess of a simple exponentially cut-off model. This likely arises from
the extended hard spectrum of the P2 component, and drives the best-fit
value for $b$ below 1. Clearly the phase-averaged spectrum
is a superposition of emission from a variety of distinct physical regions,
and we cannot expect a simple cut-off power-law spectral model to be an exact 
representation of the data.  Reduced systematic errors and the photons of 
a year or more of LAT data will be required to explore the phase-resolved 
spectral structure.

\section{Discussion}

\subsection{Light Curve} 

	The early LAT data have already greatly clarified the $\gamma$-ray
emission of the Vela pulsar and, even before detailed phase-resolved spectral
measurements and model comparisons, significantly constrain the origin of 
high energy pulsar emission. Pulsar particle acceleration
is believed to occur in the open zone, on the field lines
above each magnetic pole which extend through the light cylinder. These subtend $\sim 0.5$
of pulsar phase at high altitude, so with a total phase in the narrow
pulses of $\sim 0.06$ one would expect a comparable
fraction of the open zone, and hence fraction of the spin-down power, to
contribute to the $\gamma$-ray emission.  The narrow peaks are often 
interpreted as caustics (Morini 1983; Romani \& Yadigaroglu 1995, 
Dyks et al.\ 2004), where the combination
of field retardation, aberration and time delay causes emission from the boundary of
an acceleration zone (gap) to pile up in pulsar phase. A simple pulse
from a 1-D fold caustic would increase to a maximum as
$I(\phi) \propto (\phi-\phi_0)^{-1/2}\theta[\phi-\phi_0]$
as one approaches the destruction of an image pair at phase $\phi_0$ 
or fall off in mirror form after image pair creation.
Vela's P2 structure with a slow rise and
abrupt fall looks like a trailing (image pair destruction) caustic,
albeit with a steeper increase to the maximum, suggesting intensity variation
along the emission surface or a higher-order catastrophe. The shape of
P1 is less clear, but the apparent fast rise and slow fall would indicate a
leading (image pair creation) caustic (Figure 1).  

	The broad structure of the P3 interpulse
region suggests a beam of emission from a bundle of field lines
bracketed by the two caustic-forming zones. The similarity between
the $\gamma$-ray P3 phase structure and that of the lower frequency optical/UV
emission at $\phi \approx 0.25$ suggests a common origin -- 
they both cover similar phase intervals
and the peak shifts to later phase at higher energy.  One appealing 
possibility is to associate the optical/UV P3 with synchrotron emission 
from relatively low altitude pair cascades produced by the few-GeV curvature radiation 
dominating the $\nu F_\nu$ spectral energy distribution (SED). These
$\gamma\gamma\longrightarrow$\,e$^\pm$ pairs, with electron energy
$\gamma_e \sim 2\,{\rm GeV}/m_e c^2 \sim 10^{3.5}$, will 
produce synchrotron emission with characteristic
peak energy $\epsilon \sim 12 B_{12}\gamma_e{\rm sin}\Psi_{-1}$\,keV as they radiate
away their pitch angle $\sim 0.1\Psi_{-1}$ radians in a magnetic field of
$10^{12}$\,G. For Vela, we have $B_{12} \sim 3 (r/R_\ast)^{-3}$,
so to produce $\epsilon \sim 5$\, eV photons the pairs would reside at 
$r\sim 80 \Psi_{-1}^{-1/3}R_\ast$, where $R_\ast$ is the neutron star radius. 
If the synchrotron emission 
is outward beamed, along with the pairs, then the Compton scattered 
flux would appear at $\sim \Psi \gamma_e^2 \epsilon \sim 5$\,GeV, matching
the LAT P3 component. This suggests that the pulse of seed photons at IR to UV energies
should have energy-dependent phase shifts tracking their 
Compton-up-scattered P3 progeny.

\subsection{Luminosity and Beaming}

	While the VLBI parallax provides a well-determined 
distance of $D=287^{+19}_{-17}$\,pc, the pattern
of the $\gamma$-ray beam and hence the true $\gamma$-ray luminosity
are uncertain. To compute the true luminosity 
\begin{equation}
L_{\gamma} = 4\pi f_\Omega F_{\rm E,obs}  D^2
\end{equation}
we need a correction factor $f_\Omega$ from the
observed phase-averaged energy flux F$_{\rm E,obs}$ at the Earth line of sight
(at angle $\zeta_E$ to the rotation axis) to an average over the sky.
For a given model, the correction depends on the geometry, so we have
\begin{equation}
f_\Omega (\alpha,\zeta_E)=\int F_\gamma(\alpha,\zeta,\phi){\rm d}\Omega/(F_{exp}4\pi),
\end{equation}
where the model for a given magnetic inclination $\alpha$ from the rotation axis
gives $F_\gamma(\alpha,\zeta,\phi)$, the pulse profile as a function of phase $\phi$
seen at viewing angle $\zeta$. 
$F_{exp}$ is the expected phase-averaged flux for the actual Earth 
line-of-sight $\zeta_E$.  In general highly-aligned models, such as the polar
cap scenario, have small sky coverage and $f_\Omega<0.1$, while outer-magnetosphere models
have much larger $f_\Omega$ (Romani \& Yadigaroglu 1995). For Vela, 
the viewing angle is estimated to be
$\zeta \approx 63^\circ$ from the geometry of the {\em Chandra X-ray Observatory}-measured X-ray
torus (Ng \& Romani 2008). The magnetic impact angle $\beta=\zeta-\alpha$ is
constrained by the radio polarization data. Together,
these allow us to compute the high energy beam shape (assuming uniform
emissivity along gap surfaces). For the outer gap (OG)
picture we compute $f_\Omega=1.0$ for $\zeta=64^\circ$. For
the two-pole caustic (TPC) model appropriate to the slot gap picture,
we find $f_\Omega = 1.1$ (Watters et al.\ 2008).

The observed energy flux obtained by integrating $E_\gamma$ times the
photon number flux in Equation (1) over the 
range 0.1$-$10\,GeV is (7.87$\pm0.33\pm1.57$)$\times$10$^{-9}$\,ergs/cm$^2$/s.
For a distance of 0.287 kpc, the $\gamma$-ray luminosity of Vela is then
7.8$\times$ 10$^{34}f_\Omega$\,ergs/s.  
Thus the observed $\gamma$-ray flux implies an efficiency of
$\eta_\gamma = 0.011 f_\Omega/I_{45}$ with $f_\Omega \ge 1$ for both 
the OG and TPC models.
Note that this is a substantial fraction of the geometrical
bound on the efficiency from the peak widths noted above.

\subsection{Gamma-Ray Pulsar Models}

Three general classes of models have been discussed for $\gamma$-ray pulsars.  
In polar cap models (Daugherty \& Harding 1994, 1996; 
Sturner \& Dermer 1995; Harding \& Muslimov 2004), the particle acceleration and 
$\gamma$-ray production takes place in the open field line region within 
one stellar radius of the magnetic pole of the neutron star.  In outer gap models (Cheng,
Ho \& Ruderman 1986; Romani 1996;
Hirotani 2005), the interaction region lies in the outer magnetosphere in vacuum gaps 
associated with the last open field line. Other recent models hark back to the `slot-gap'
picture of Arons (1983), with the polar cap rim acceleration extending 
to many stellar radii (Harding \& Muslimov 2004).

The appreciable offset from the radio peak, coupled with the presence 
of only one radio pulse, suggests that the $\gamma$-rays
arise at high altitude. This is also supported by the lack of hyper-exponential
absorption in the spectrum, which is a signature of $\gamma$-B pair
attenuation of low altitude emission. The remaining two geometrical models can 
produce the general double peak profile of Vela (and many other $\gamma$-ray pulsars).
In `Outer gap' models emission starts near the `null charge'
$\Omega \cdot B=0$ surface at radius $r_{\rm NC}$ and extends toward the 
light cylinder $r_{\rm LC} = cP/2\pi$. One magnetic
pole dominates the emission in each hemisphere and the two peaks represent
leading and trailing edges of the hollow cone of emission from this pole.
If on the other hand emission extends well below the null charge surface
toward the neutron star, both magnetic poles can contribute toward emission
in a given hemisphere. This is the two-pole caustic model of Dyks \& Rudak (2003) 
which might be realized in `slot gap' acceleration models. In this case the
leading `image pair creation' caustic from high altitudes should not be visible,
and the first $\gamma$-ray pulse represents the trailing caustic 
from emission at $r<r_{NC}$ 
below the null-charge surface (the radio magnetic pole) while the second
pulse represents a trailing caustic at higher altitudes from the opposite pole; 
both pulses should generally show a slow rise and steeper fall.  
Detailed $\gamma$-ray pulse profiles and, 
especially, phase resolved spectra can help distinguish these scenarios.

Another geometric difference in the two scenarios is the origin of the
emission outside the
main pulse. Low altitude field lines in the two-pole model tend to produce emission
along all lines of sight. In the outer gap picture high altitude field lines
can contribute faint emission extending through the off pulse region.

Further, as noted above, the energy spectrum also presents challenges
for near-surface emission.  In the polar cap models, a sharp turnover 
is expected in the few to 10\,GeV energy range due to attenuation 
of the $\gamma$-ray flux in the magnetic field (Daugherty \& Harding 1996).  
The spectral change at $\sim$2.9\,GeV does not appear to fit this model.
We conclude that low altitude radiation subject to $\gamma$-B pair
production cannot account for the bulk of the Vela $\gamma$-ray emission.
Indeed we can use the observed cut-off to estimate a minimum emission
height as 
$r\approx (\epsilon_{max} B_{12}/1.76{\rm GeV})^{2/7}P^{-1/7} R_\ast$,
where $\epsilon_{max}$ is the unabsorbed photon energy, $P$ is the
spin period and the surface field is $10^{12}B_{12}$\,G (Baring 2004). 
Using Vela parameters ($P=0.089$\,s, $B_{12}=3.4$) we see that our cut-off
implies that the bulk of the emission arises from $>2.2R_\ast$. Since we see pulsed photons
up to $\epsilon_{max} =17$\,GeV, this emission must arise at $r>3.8R_\ast$.
A similar conclusion has been recently made for the Crab pulsar
by the MAGIC team, who observe pulsed $E>25$\,GeV emission, using
the imaging air Cerenkov technique (Aliu et al.\ 2008).    % Science, submitted).
Finally, it should be noted that while typical radio pulsar emission
is inferred to arise at relatively low altitudes, for Vela-type
pulsars recent radio models infer emission heights of as much as 
$100R_\ast$ (Karastergiou \& Johnston 2007). All of these factors
point to a high-altitude origin of the Vela $\gamma$-ray emission.
\bigskip
\bigskip
\bigskip
\bigskip

\section{Summary}

The early LAT observations of Vela serve to show that the instrument is 
performing very well, with effective area comparable to expectations, 
good PSFs and excellent source localization, especially at high energies. 
The time-tagging of events is excellent and our 
pulse profile reconstruction is limited primarily by the accuracy of
the radio-derived pulsar ephemeris. This presently has an accuracy of
$\sim$100\,$\mu$s over our analyzed data span. The energy response and
effective area calibration are currently being validated, but
are at present known to $\sim 5$\% near 1\,GeV, with uncertainties increasing
to $\sim 20$\% at $\le0.1$\,GeV and $\sim 30$\% at $\ge 10$\,GeV.

These data also substantially improve our knowledge of the pulse properties of
the Vela pulsar (PSR B0833$-$45) and are now placing important constraints on
theoretical models.  Already in these data we see: 

1. The pulse profile is complex, with P1 and P2 dominated by very narrow components
and substantial structure in the `bridge' region. 

2. Although the P1 and P2 phases are very stable across the $\gamma$-ray band, 
the the P1/P2 ratio decreases with energy. There is a distinct third
peak in the `bridge' component which sharpens and moves to later pulsar phase
with increasing energy.

3. While faint emission appears within $\sim 1^\circ$ of the pulsar
in the off-pulse phase, association if any with Vela is still unclear,
due to the bright Galactic background in the vicinity.
We thus quote an upper limit on the unpulsed point source flux, 
at 2.8\% of the phase-averaged pulsed flux.
A true PWN association will require additional data,
allowing resolution at higher energies and a match to the 
radio and TeV images. 

4. The phase-averaged $\gamma$-ray energy spectrum can be represented by a power law, 
with a exponential cut-off at $2.9\pm 0.1$ GeV. The hyper-exponential cut-off 
index $b = 0.88\pm 0.04_{-0.52}^{+0.24}$ is not significantly different from 
the simple exponential 
value $b=1$. Large values of $b$, as expected for models radiating from the 
near-surface polar cap zone, are excluded.

	During the first-year sky survey, we expect to collect 
approximately $1.3 \times 10^5$ additional pulsed Vela photons. This will allow a 
substantial increase in sensitivity to narrow pulse components. We also expect
improvements in our understanding of the instrument that will allow us to
significantly reduce systematic errors. These two improvements
will allow derivation of high accuracy spectral 
properties in more than 100 bins of pulsar phase. These are the observational results
that should let us pin down the location and energy distribution of the 
radiating particles. In turn this should help us `reverse-engineer'
this $\gamma$-ray machine and should provide substantial insight into the physics
of pulsar magnetospheres.

\acknowledgments
The GLAST LAT Collaboration acknowledges the generous
support of a number of agencies and institutes that have
supported the development of the LAT. These include the
National Aeronautics and Space Administration and the Department
of Energy in the United States, the Commissariat \`a
l'Energie Atomique and the Centre National de la Recherche
Scientifique / Institut National de Physique Nucl\'eaire et de
Physique des Particules in France, the Agenzia Spaziale Italiana,
the Istituto Nazionale di Fisica Nucleare, and the Istituto
Nazionale di Astrofisica in Italy, the Ministry of Education,
Culture, Sports, Science and Technology (MEXT), High
Energy Accelerator Research Organization (KEK) and Japan
Aerospace Exploration Agency (JAXA) in Japan, and the K.
A. Wallenberg Foundation and the Swedish National Space
Board in Sweden.

The Australia Telescope is funded by the Commonwealth of
Australia for operation as a National Facility managed by the CSIRO.

%\vfil
%\eject

\vfill\eject

%\end{document}

%%%%%%%%%%%%%%%%%%%%%%%%%%%%%%%%%%%%%%%%%%%%%%%%%%%%%%5
\placetable{2}
Data table for Electronic version.
\begin{table*}
\centering
\caption{\label{2} Phase Averaged Spectral Points for the Vela Pulsar}
\bigskip
\begin{tabular}{lcccc}
\hline\hline
Energy (GeV) & Differential flux F (erg cm$^{-2}$s$^{-1})$ & 
$\Delta$F$_{stat}$ & $\Delta$F$_{syst}$ & 
$\Delta$F$_{tot}$\footnote{$\Delta$F$_{syst}$ is the systematic error on 
the flux and $\Delta$F$_{syst}$ are the statistical errors on the flux, 
while $\Delta$F$_{tot}$ is the total error on the flux}\\
\hline
0.14	&	1.381E-09	&	5.805E-11	&	2.952E-10	&	3.009E-10	\\
0.18	&	1.385E-09	&	5.191E-11	&	1.601E-10	&	1.683E-10	\\
0.22	&	1.514E-09	&	3.895E-11	&	2.261E-10	&	2.294E-10	\\
0.28	&	1.734E-09	&	4.212E-11	&	2.301E-10	&	2.340E-10	\\
0.35	&	1.953E-09	&	4.563E-11	&	2.232E-10	&	2.279E-10	\\
0.45	&	2.129E-09	&	5.066E-11	&	2.625E-10	&	2.674E-10	\\
0.56	&	2.088E-09	&	5.289E-11	&	1.671E-10	&	1.752E-10	\\
0.71	&	2.200E-09	&	5.711E-11	&	2.755E-10	&	2.814E-10	\\
0.89	&	2.405E-09	&	6.505E-11	&	3.006E-10	&	3.076E-10	\\
1.12	&	2.417E-09	&	7.111E-11	&	2.185E-10	&	2.298E-10	\\
1.41	&	2.493E-09	&	8.007E-11	&	4.626E-10	&	4.695E-10	\\
1.78	&	2.312E-09	&	8.507E-11	&	4.101E-10	&	4.188E-10	\\
2.24	&	2.273E-09	&	9.770E-11	&	4.164E-10	&	4.277E-10	\\
2.82	&	2.005E-09	&	1.055E-10	&	3.896E-10	&	4.036E-10	\\
3.55	&	1.617E-09	&	1.052E-10	&	4.524E-10	&	4.645E-10	\\
4.47	&	1.357E-09	&	1.135E-10	&	4.286E-10	&	4.434E-10	\\
5.62	&	1.019E-09	&	1.116E-10	&	1.854E-10	&	2.164E-10	\\
7.08	&	4.403E-10	&	8.276E-11	&	1.931E-10	&	2.101E-10	\\
8.91	&	4.435E-10	&	9.914E-11	&	1.870E-10	&	2.116E-10	\\
11.22	&	9.390E-11	&	6.150E-11	&	0.000E+00	&	2.542E-11	\\
14.13	&	4.191E-11	&	4.550E-11	&	0.000E+00	&	3.043E-11	\\
17.78	&	1.169E-10	&	8.417E-11	&	1.053E-11	&	8.483E-11	\\
22.39	&	0.000E+00	&	0.000E+00	&	0.000E+00	&	0.000E+00	\\
28.18	&	0.000E+00	&	0.000E+00	&	0.000E+00	&	0.000E+00	\\
\hline
\end{tabular}
\end{table*}
%%%%%%%%%%%%%%%%%%%%%%%%%%%%%%%%%%%%%%%%%%%%%%%%%%%%%%%%%%%%%%%%%
\end{document}